\documentclass[aps,prl,twocolumn,superscriptaddress,showpacs,showkeys,amsmath,amssymb]{revtex4}
\usepackage{graphicx}
\usepackage{dcolumn}
\usepackage{bm}
\usepackage{color}
\usepackage[thinspace,thinqspace,amssymb,pstricks]{SIunits}
\usepackage{longtable}
\usepackage[hypertex]{hyperref}
\begin{document}

\title{Resonant Photoelectron Diffraction with circularly polarized light}

\date{\today}

\author{Martin Morscher}
\affiliation{Physik-Institut, Universit\"{a}t Z\"{u}rich, Winterthurerstrasse 190, CH-8057
Z\"{u}rich, Switzerland}

\author{Frithjof Nolting}
\affiliation{Paul Scherrer Institut, CH-5232 Villigen PSI, Switzerland}

\author{Thomas Brugger}
\affiliation{Physik-Institut, Universit\"{a}t Z\"{u}rich, Winterthurerstrasse 190, CH-8057
Z\"{u}rich, Switzerland}

\author{Thomas Greber}
\email{greber@physik.uzh.ch}
\affiliation{Physik-Institut, Universit\"{a}t Z\"{u}rich, Winterthurerstrasse 190, CH-8057
Z\"{u}rich, Switzerland}

\begin{abstract}
Resonant angle scanned x-ray photoelectron diffraction (RXPD) allows the determination of the atomic \emph{and} magnetic structure of surfaces and interfaces.   
For the case of magnetized nickel the resonant $L_2$ excitation with circularly polarized light yields electrons with a dichroic signature from which the dipolar part may be retrieved.
The corresponding $L_2MM$ and $L_3MM$ Auger electrons carry different angular momenta since their source waves rotate the dichroic dipole in the electron emission patterns by distinct angles.

\end{abstract}



\maketitle

The quest for atomic scale structure determination at surfaces and interfaces lead to the development of a large number of powerful methods \cite{vho99}. Among those,  x-ray photoelectron spectroscopy (XPS) with angular resolution (XPD) allows structure determination paired with chemical and magnetic sensitivity \cite{fad10}. 
The signal is best, when the x-ray absorption coefficient is at maximum.
These maxima occur in resonant excitation and have so far been exploited for the probing of defect states in TiO$_2$ \cite {Kruger2008}, for looking inside an endofullerene \cite{Treier2009}, or for the investigation of the mixed valence structure of magnetite \cite{mag10}.

In this letter, angle scanned resonant photoelectron diffraction (RXPD) is applied to nickel, the "fruit fly" of resonant photoemission \cite{Guillot1977,fel79,tjeng,vde94,Weinelt1997}. 
Circularly polarized light is used for the precise measurement of the dipole induced by the x-ray magnetic circular dichroism which is largest at resonance.
After this proof of principle we show that the angular momentum of the outgoing electrons can be measured with RXPD.  This extends the results of Daimon et al., who demonstrated that forward scattering XPD patterns rotate due to the angular momentum of circularly polarized photons \cite{dai93,daimon2001,matsui2010}.
Here, the electron source wave \cite{gre91,gre92} rotates the magnetisation direction in the final state of the emitted electrons.
It is also found that Auger electrons may carry angular momenta opposite to that of the exciting photon and larger  than $\hbar$.

Circular magnetic dichroism $\Delta I_{MD}$ is the difference between the absorption coefficient of right and left circularly polarized light.  It is proportional to the scalar product of the magnetization $\bf m$ and the angular momentum of the incoming photon ${\bf L}_{ph}$ \cite{Stohr2006}.
\begin{equation}
\Delta I_{MD} \propto {\bf m} \cdot  {\bf L}_{ph}
\label{E0}
\end {equation}
For right circularly polarized light ($\sigma^+$) ${\bf L}_{ph}$ is parallel to the propagation direction of the photon, and for left circularly polarized light ( $\sigma^-$)  ${\bf L}_{ph}$ is antiparallel.
$\Delta I_{MD}$ is a dipole, i.e. proportional to $\cos(\vartheta)$, where $\vartheta$ is the angle between $\bf m$  and  ${\bf L}_{ph}$. From this follows that the absolute orientation of ${\bf m}$ is determined from three or more non-coplanar light incidences.

Photoemission yields $\Delta I_{MD}$, under the assumption that the photoemission current is proportional to the x-ray absorption coefficient.
This must not hold for partial measurements like in angular resolved photoemission where $dI/d\Omega$ is measured \cite{westphal1989}. 
The different photoelectron source waves of electrons excited with differently polarized light lead to different photoemission final states that comprise information on the  magnetism {\emph {and}} the surrounding of the emitter \cite{chasse2005}. 
Here we show that contributions of the atomic and the magnetic structure can be disentangled with a spin integrated experiment. 

\begin{figure}
\center
\includegraphics[width=0.9\columnwidth]{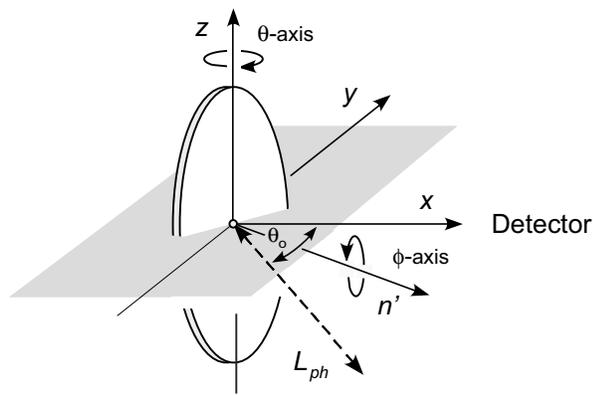} 
\caption{Geometry of the XPD experiment. The electron detection is parallel to $\bf x$ and the polar ($\theta$)  rotation axis parallel to $\bf z$. The sample normal ${\bf n}'$ is the azimuthal ($\phi$) rotation axis and lies with the light incidence along its angular momentum ${\bf L}_{ph}$ in the $xy$ plane (shaded area), $\theta_o$ =55$^\circ$ away from ${\bf x}$.}
\label{F1}
\end{figure}

The experiments have been performed at the SIM beamline at the Swiss Light Source
(SLS) \cite{papersim} in an endstation dedicated for x-ray photoelectron spectroscopy
(XPS) and angle-resolved x-ray photoelectron diffraction (XPD) with a base pressure below $2\cdot 10^{-10}$ mbar. 
The x-rays impinge perpendicular to the polar rotation axis with an angle $\theta_o$ between the x-rays and the electron energy analyzer of 55$^\circ$ (see Figure \ref{F1}).
The degree of polarization is better than 98\%.
All measurements are done at room temperature.
The Ni(111) yoke crystal \cite{oku09} was
cleaned by repeated cycles of argon sputtering and annealing. 
It is magnetized by passing a current of 2 A for 30 s through the yoke coil. The resulting magnetization was inferred from an x-ray magnetic circular dichroism (XMCD) spectrum at the Ni L$_{2}$,L$_{3}$-edges in the total-electron-yield mode. Comparison with corresponding spectra of Chen et al. \cite{Chen1991} indicate a magnetization of $\sim$40\%, which is not 100 \% due to a multidomain structure.

\begin{figure}
\includegraphics[width=1\columnwidth]{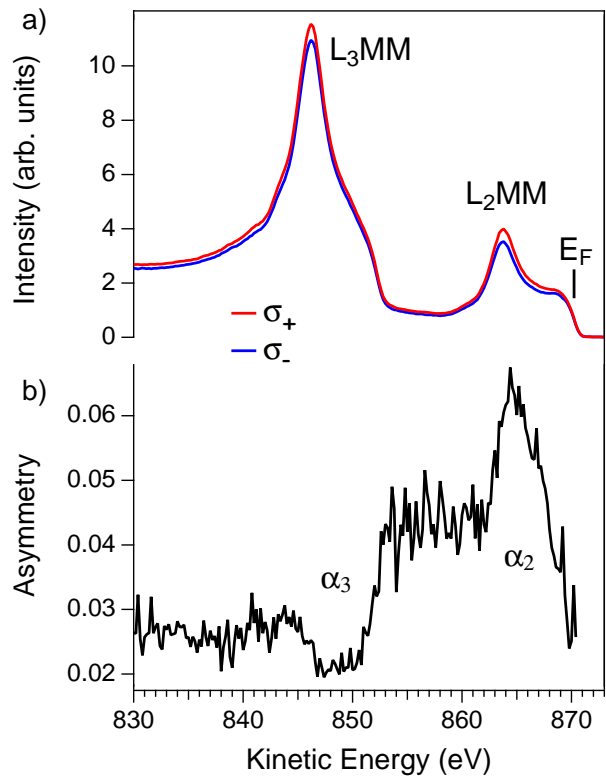}
\caption{(Color online) (a) Near normal photoelectron spectra with left and right circularly polarized light with an angle of 30$^\circ$ and 180-30$^\circ$ between ${\bf L}_{ph}$ and $\bf m'$. The photon energy is set on the Ni L$_2$-resonance at $\hbar \omega=870.5$ eV. 
The spectra have been normalized with the photon flux.
The Fermi level $E_F$, the $L_2MM$ and the $L_3MM$ Auger deexcitation peaks are indicated.
(b) The asymmetry $A$ between right and left circularly polarized light exhibits two distinct extrema $\alpha_2$ and $\alpha_3$, on which we performed XPD measurements.}
\label{F2}
\end{figure}

Resonant photoemission on 3d transition metals is most intense at the L$_3$ absorption edge \cite{Chen1991,Kruger2008}.
Here we investigate the L$_2$ resonance since it provides L$_2$MM {\it{and}} L$_3$MM emission which allows for direct comparison and consistency checks. 
Figure \ref{F2} shows resonant x-ray photoelectron spectra from magnetized Ni(111) of right and left circularly polarized light. 
The photon energy is set on the Ni L$_{2}$-resonance (2p$_{1/2}\rightarrow$3d) at $\hbar \omega=870.5$ eV. 
The Fermi level $E_F$ at 870.5 eV electron energy, the $L_2MM$ (863.8 eV) 6 eV satellite (see \cite{Huefner} and references therein), and the $L_3MM$ (846.2 eV) Auger deexcitation peak are most prominent.
The spectra have been normalized with the photon flux.
Figure \ref{F2}b) demonstrates circular dichroism in these resonantly excited electron emission spectra.
The asymmetry $A=(I(\sigma^+)-I(\sigma^-$))/($I(\sigma^+)+I(\sigma^-))$ between right and left circularly polarized light exhibits a maximum at $\alpha_2$ and a minimum at $\alpha_3$. 
The asymmetry can be reversed by switching the magnetization or by the rotation of the sample by 180$^\circ$ \cite{mor11}.
Off resonance, at  $\hbar \omega=873.5$ eV the dichroic asymmetry in the  $L_{2}MM$ Auger line is 
1.4 $\pm$ 0.8 \% (data not shown).
The extrema $\alpha_2$ at 848.4 eV and $\alpha_3$ at 863.6 eV do not exactly coincide with the $L_2MM$ and the $L_3MM$ intensity maxima, which indicates multiplet structure \cite{magnuson1998}.
In the following we use the labels $\alpha_i^\sigma$ for electrons at the energies of $\alpha_2$ and $\alpha_3$, excited with $\sigma^+$ and $\sigma^-$ polarized radiation, respectively.

If we perform on these resonances angle scanned x-ray photoelectron diffraction (RXPD), the experiment yields information on the atomic {\emph {and}} the magnetic structure.
Briefly, the sample frame $(x',y',z')$ is rotated in the lab frame $(x,y,z)$. The photoelectron intensity $I$ is mapped in polar coordinates $(f(\theta), \phi)$, where the polar angle $\theta$ and the azimuthal angle $\phi$ define the sample orientation with respect to the electron detection direction 
(see Figure \ref{F1}) \cite{osterwalder1991}.
This leads for $\Delta I_{MD}$ to a dipolar function  $D(\theta,\phi)$ in the XPD map  that depends on ${\bf L}_{ph}$, the electron detection direction in the lab frame and ${\bf m'}(\theta_m,\phi_m, a_m)$ in the sample frame, where the amplitude $a_m$ is a measure for the magnitude of the dichroism  \cite{mor11}.

\begin{figure}
\includegraphics[width=1\columnwidth]{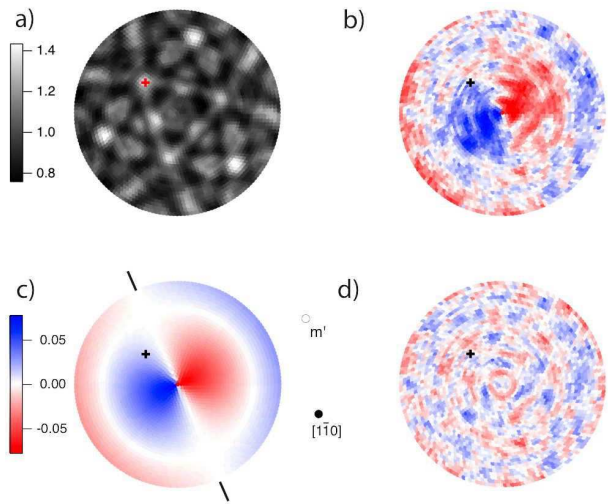}
\caption{(Color online) (a) Resonant x-ray photoelectron diffraction (RXPD) data of Ni(111). The  3500 stereographically projected data points for polar angles $0\leq\theta\leq70^\circ$ of the intensity at $\alpha_2$ ($E_{kin}$=863.8 eV,  $\hbar\omega$ = 870.5 eV, $ \sigma^+$)  in Figure \ref{F2}. 
The data are $\phi$-averaged, i.e. normalized at each polar angle with the corresponding average intensity.
(b) Asymmetry of two XPD data sets at $\alpha_2$  measured with right and left circularly polarized light. The twofold i.e. dipolar pattern reveals the direction of the magnetization. 
(c) $D$-function-fit to the asymmetry in (b). The direction of magnetisation ${\bf m'}$ and ticks for the corresponding azimuthal orientation at $\phi_m \pm90^\circ$ are indicated.
The [$011$]-direction (cross), and the yoke axis along [$1\bar{1}0$] are marked.
(d) Residuum of the fit in (c) with respect to (b).
}
\label{F3}
\end{figure}

Figure \ref{F3}a) shows RXPD data for  $\alpha_2^+$.
The XPD map is dominated by the information on the atomic structure which corresponds to that of a face centered cubic $(fcc)$ crystal which is cut along the (111) plane  \cite{wider2001}.
Below the obvious atomic structure dichroic information must be hidden.
In order to visualize the dichroism, we form the asymmetry $A$ between the $\alpha_2^+$ and the $\alpha_2^-$ XPD scans (see Figure \ref{F3}b).
These data contain information on the dipolar (magnetic) nature of dichroism and higher order multipoles, which are related to differences in the diffraction patterns due to different source waves \cite{dai93,gre01}.
The dipolar part has the symmetry  $A(\theta,\phi)=-A(\theta,\phi+180^\circ)$, as it is expected for in plane magnetisation. 
Figure \ref{F3}c) shows the fit of a dipolar function  $D(\theta,\phi)$, which determines ${\bf m'}$. 
We find $\theta_m=89.0 \pm 1^\circ$, and $\phi_m=39.1 \pm 1^\circ$, where $\phi=0$ is set to the $[\bar{1}10]$ direction.
This result is consistent with spin polarized photoemission \cite{oku09}, though the magnetization is not aligned along the second easy axis $[1\bar{1}0]$ as it was the case for an other Ni(111) picture frame crystal \cite {Donath1994}.
The rotation of the sample and the incidence of the light  impose on $D$ {\emph{two}} nodal lines (${\bf m'}\cdot {\bf L}_{ph}=0$): A circle at $\theta_o=55^\circ$, and a diameter perpendicular to $\phi_m$.
In Figure \ref{F3}d) the residuum of the asymmetry and $D$ is shown. 
It has the C$_3$ symmetry of the substrate and indicates further differences in the diffraction patterns due to different source waves created by $\sigma^+$ and $\sigma^-$ photons, respectively. 
Such effects have been pioneered by Daimon et al., where they showed that the angular momenta of the photons are transferred to the photoelectrons, which in turn lead to an emitter scatterer distance dependent rotation of the forward scattering peak \cite{dai93,daimon2001}.

For the data in Figure \ref{F3} b) we expect no photon induced rotation because in the asymmetry between $\sigma^+$ and $\sigma^- $any such effect should be canceled.
This changes, when a dipolar function $D$ is fitted to an individual XPD scan with either $\sigma^+$ or $\sigma^-$ radiation.
If we fit a dipolar function $D$ to the data in Figure \ref{F3}a), or the ones recorded with $\sigma^-$ polarization, we find magnetisation directions which are within $\pm 6^\circ$ consistent with the magnetization direction as found from Figure \ref{F3} b).
Although this has the practical advantage that the magnetization is determined without switching the light polarization, it is not very accurate since the $D$-function is much weaker than the forward scattering induced XPD patterns.

If we want to extract more quantitative information on the rotation of the $D$-functions upon use of photons with plus or minus $\hbar$ angular momentum we have to perform a normalisation that removes the forward scattering intensity modulations but preserves, in contrast to the asymmetry used in Figure \ref{F3}  the angular momentum of $\sigma^+$ or  $\sigma^-$.
We do so in using the $\phi$-averaged data ${\bar{\alpha}}_i^\sigma$ and form $\Delta{\alpha_2^\sigma}=2\cdot {\bar{ \alpha}}_2^\sigma /  ({\bar{\alpha}}_3^+ + {\bar{\alpha}}_3^-)-1$ and vice versa.

As $\alpha_2$ and $\alpha_3$ electrons are expected to have very similar XPD patterns \cite{gre92} - the wavelength-difference  between the two selected electron energies of $\alpha_2$ and $\alpha_3$ is 1\% - most XPD information on the atomic structure should be cancelled, though the $\Delta \alpha_i^\sigma$ are expected to show a polarization dependent  rotation $\Delta\phi_m(\alpha_i,\sigma)$ of the observed magnetisation direction.
For the 6 eV satellite, i.e. $L_2$MM transition,  we find a rotation $\Delta \phi_m$ of $\pm 4.2^\circ$ around the value of Figure \ref{F3}b). 
It is related to the Daimon effect, i.e.  forward scattering peak rotation \cite{dai93}. 
Essentially, the angular momentum of an outgoing photoelectron induces a rotation of all features in the XPD patterns with respect to the crystal lattice.
For single scattering the maximum angle of rotation $\gamma_{max}$ is given by $n\cdot \hbar/(R \cdot p \cdot \sin^2(\theta_o))$, where $R$ is the distance between emitter and scatterer, $p$ the momentum of the outgoing electron and $\theta_o$ the angle between the light incidence and the electron detection (see Figure \ref{F1}) \cite{daimon2001,chasse1997}. 
For nickel, $n=1$, an electron kinetic energy of 850 eV and $\theta_o$=55$^\circ$, $\gamma_{max}$ gets $2.2^\circ$.
Of course, the angular shift is not isotropic, it depends on the angle between the electron angular momentum and the nearest neighbour directions.  However, for an $fcc$ material as it is nickel, the 12 nearest neighbours of an emitter in the bulk sit on a sphere with radius $R$, on the vertices of a cuboctahedron and must lead to a fairly isotropic rotation of the XPD patterns around the axis of the incoming photons.
The $\Delta \phi_m ( \alpha_2)$'s  have the same sense of rotation as the corresponding photon angular momentum and are compatible with the transfer of 2$\hbar$ of angular momentum to the emitted electrons.

\begin{figure}
\includegraphics[width=0.95\columnwidth]{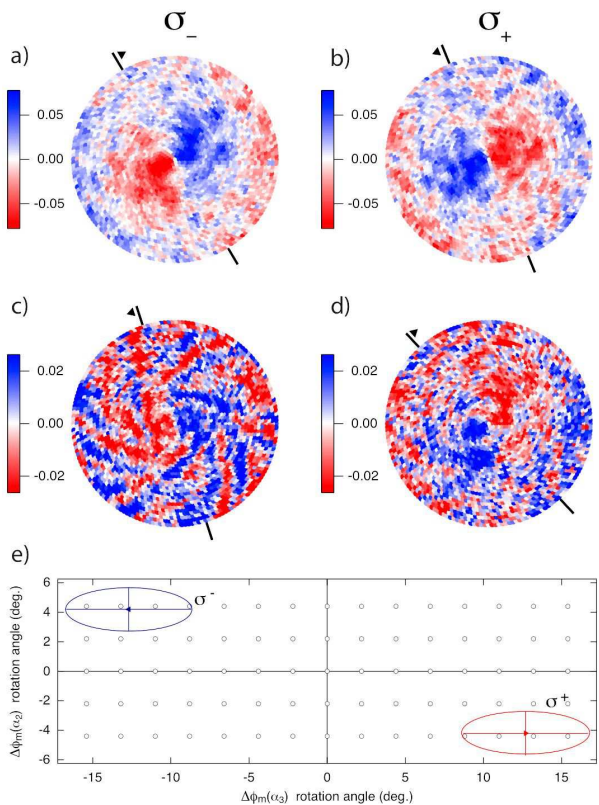}
\caption{(Color online) (a)-(d): Individual XPD patterns with the same orientation as in Figure \ref{F3} of (a) $\Delta{\alpha_{2}}^{-}$, (b) $\Delta{\alpha_{2}}^{+}$, (c) $\Delta{\alpha_{3}}^{-}$ and (d) $\Delta{\alpha_{3}}^+$ (for definition see text). 
The ticks lie on the azimuth of the node of the corresponding dichroic dipole. 
The solid triangles indicate the rotation towards the magnetisation direction. (e) Rotation angles $\Delta\phi_m$($\alpha_{2}^\pm$) versus  $\Delta\phi_m$($\alpha_{3}^\pm$). For a given polarization the electrons rotate in opposite directions. The ellipses represent the error bars. The open circles are the rotation angles as expected from quantized angular momenta $n\cdot\hbar$. }
\label{F4}
\end{figure}

Figure \ref{F4} shows the $\Delta{\alpha_{2}^\sigma}$ and the $\Delta{\alpha_{3}^\sigma}$  XPD scans for $\sigma^+$ and $\sigma^-$ radiation.
Dipolar functions ($D$) as shown in Figure \ref{F3}c) appear, where the sign changes upon change of the polarization.
The $D$-functions  for $\Delta{\alpha_{2}^\pm}$ and $\Delta{\alpha_{3}^\pm}$ indicate $\Delta\phi_m $'s of $\pm 4.2 ^\circ$ and $\mp 12.6 ^\circ$, respectively.
We want to note that the use of more than 3000 different photon incidence angles allows a very accurate $\pm 0.8^\circ$ determination of the $\Delta\phi_m(\alpha_2)$'s  and permits for single quantum assignments 2$\hbar$ (-2$\hbar$).
In the $\Delta \alpha_3$ patterns with a lower asymmetry (see Figure \ref{F2}) the error increases by a factor of 3 and makes it compatible with angular momenta of $-6\pm2\hbar$ or $6\pm2\hbar$ (see Figure \ref{F4} e)).
This surprising result implies that a $L_3MM$ Auger electron channel produces electrons with large, opposite angular momentum compared to that of the photons.   
These results emphasise that the information on the angular momentum of an electron source wave may not only be accessed by a forward scattering peak rotation \cite{dai93}, but for magnetic systems also by the precise measurement of the  source wave dependent circular magnetic dichroism.
 
In summary it has been shown that resonant x-ray photoelectron diffraction (RXPD) is suited to extract the atomic and magnetic structure of surfaces and interfaces.
Furthermore it is demonstrated that the method directly accesses the angular momenta of the emitted electrons.

Fruitful discussions with J. Osterwalder and the support of the Swiss National Science Foundation are gratefully acknowledged. The experiments have been performed at the Swiss Light Source.
 

\end{document}